\documentstyle[sprocl]{article}
\input{epsf}
\def\be{\begin{equation}}
\def\ee{\end{equation}}
\def\bea{\begin{eqnarray}}
\def\eea{\end{eqnarray}}

\begin{document}
\title{PRELIMINARY RESULTS FROM A SEARCH FOR DISORIENTED CHIRAL CONDENSATES AT 
MINIMAX}
\author{J.STREETS}
\address{ Fermi National Accelerator Laboratory \\
Batavia, IL 60510, U.S.A}
\author{Representing the MiniMax Collaboration}
\address{ }


\maketitle\abstracts{
We report on the progress of a search for disoriented chiral
condensates (DCCs) in the far forward region at $\sqrt{s}=1.8~TeV$. 
MiniMax is a small collider experiment
situated at the C0 interaction region of the Tevatron and has been designed to
measure the ratio of charged to neutral pions produced at $\eta\approx4.1$.
The distribution of this ratio is expected to be very different
for the generic binomial and DCC particle production models.
We present a method used for the search using the factorial moments of
particle generating functions, which can be extracted directly from the data.
A preliminary comparison of robust ratios of measured values and those 
expected from the different models is also presented.}
\section{Disoriented Chiral Condensates}
There is a class of events seen in cosmic ray experiments which occur at
much higher rates than expected from studies of standard Monte Carlo
simulations.\cite{raja}
These events have high multiplicity in the forward direction 
($3.5\leq\eta\leq4.5$), abnormal ratios of electromagnetic
to hadronic energy and $\sqrt{s}\tilde{>}1~TeV$.
It has been proposed that these events can arise from the 
strong-interaction with unconventional orientation of the chiral order
parameter. This disoriented chiral condensate is then supposed to decay into
a coherent semiclassical pion field having the same chiral 
orientation.\cite{ansl}


In the DCC model, the fraction of neutral pions in the event ($f={n\over{N}}$)
can be shown to occur with a probability 
distribution $1\over{2\sqrt{f}}$ in the limit of large numbers.
The striking differences in the distributions at $f=0$ and $f=1$ for the
DCC and generic binomial models can be seen in figure 1.

\section{The MiniMax Detector}
Figure 2 shows the layout of the detector.\cite{kenk} Charged tracks are measured by
24 MWPCs which cover 
a circle of radius approximately  0.65 units in $(\eta,\phi)$ space
centered on $\eta=4.1$.
Neutral tracks are reconstructed from conversions in
a moveable lead plate, typically $1 X_0$ in thickness.
The trigger is comprised of scintillating counters placed both in the 
telescope and at $\eta=-3.5$ in the anti-proton direction.
This trigger was measured to be sensitive to $35-40~mb$ of the total
$p\bar{p}$ cross-section.

\begin{minipage}[t]{1.5in}
{\epsfxsize=1.5in \epsffile{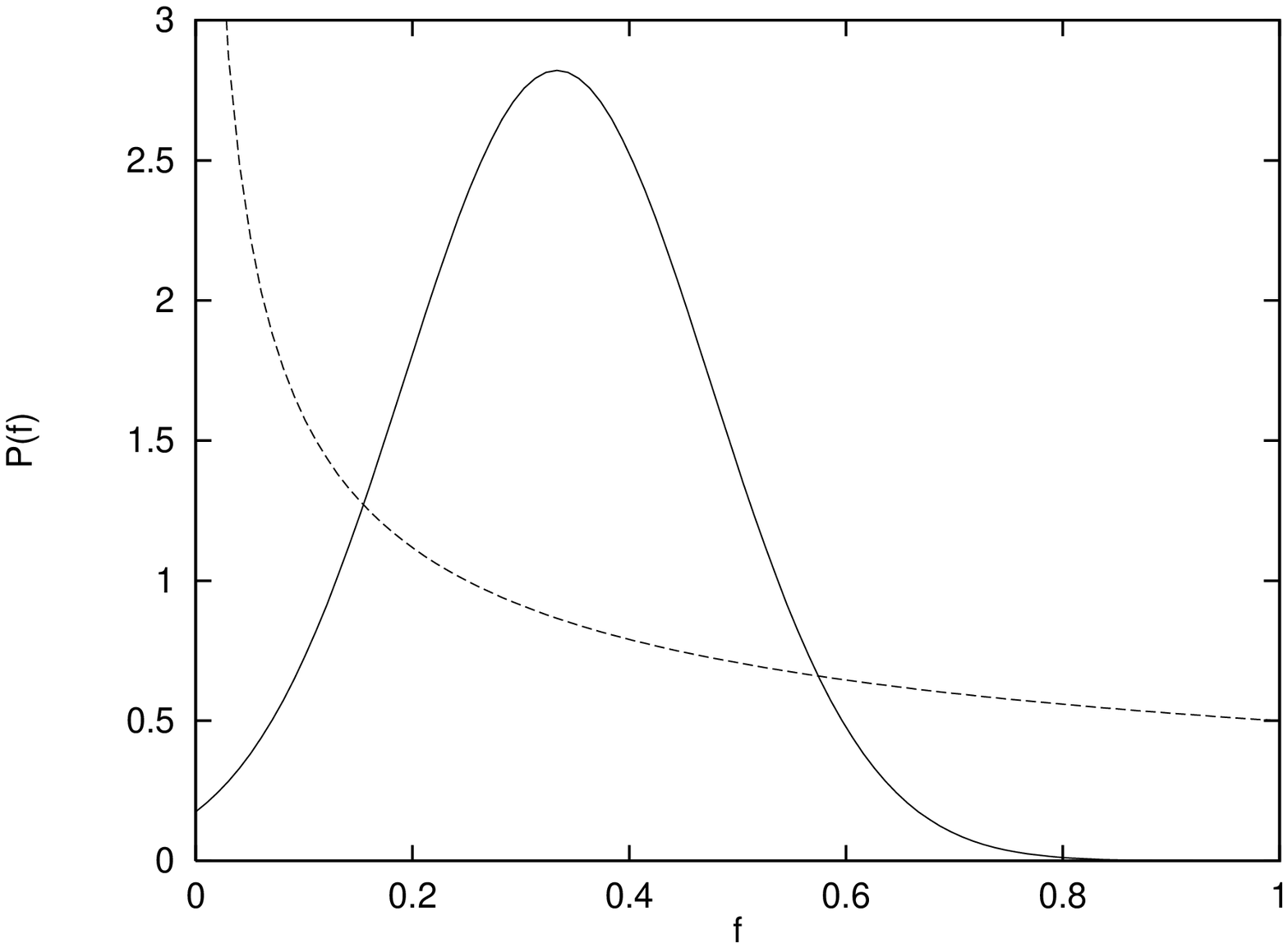} }
\small{ Figure~1:~P(f)~for~the~generic (binomial curve) and DCC models ($\propto 1/\sqrt{f}$) }
\end{minipage} \ \
\begin{minipage}[t]{2.5in}
\epsfxsize=2.5in \epsffile{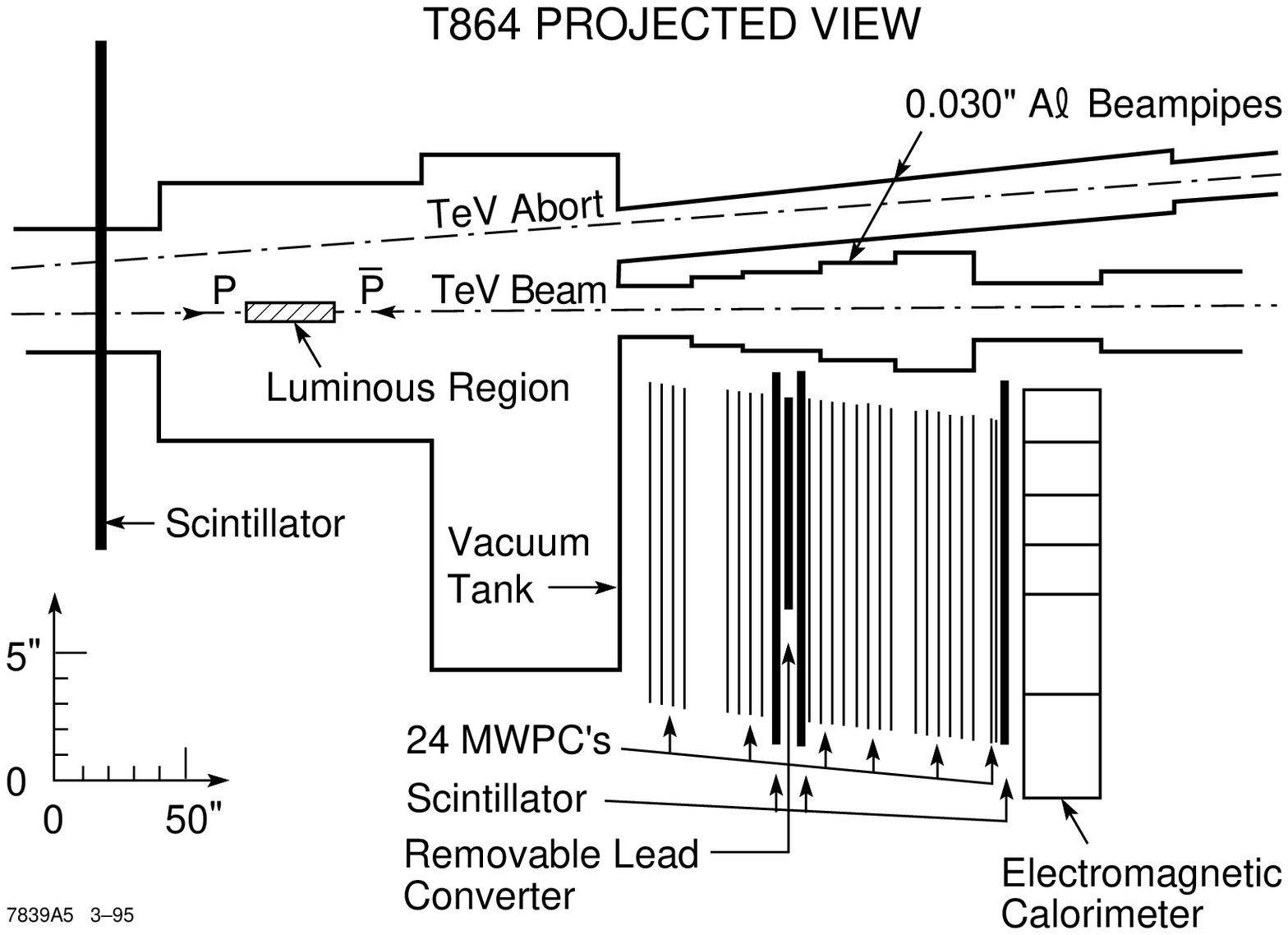}
\begin{center}
\small{ Figure 2: The MiniMax Detector}
\end{center}
\end{minipage}
\section{Using Moments of the Generating Functions}
The factorial moments of the generating functions for bivariate distributions
of $m$ charged pions and $n$ observed gamma rays are expressed as $f_{mn}$.
The first few moments reduce to robust observables if we assume that the
detection efficiencies of tracks are independent of track multiplicity.
\begin{equation}
\begin{array}{ll}
f_{10} = <N_{ch}> & f_{20} = <N_{ch}(N_{ch}-1)>  \\
f_{01} = <N_{\gamma}> & f_{21} = <N_{ch}(N_{ch}-1)N_{\gamma}> \\
f_{11} = <N_{ch}N_{\gamma}> & f_{30} = <N_{ch}(N_{ch}-1)(N_{ch}-2)>
\end{array}
\end{equation}
The ratios $r_{m1} = {{f_{m1}f_{10}}\over{f_{(m+1)0}f_{01}}}$ all yield
$1$ for generic distribtions and $1\over{m}$ for pure DCC.
They have many nice properties.
In particular, they are independent of the form of the parent pion 
multiplicity distribution,
of the (uncorrelated) detection efficiency for charged tracks, of the 
(uncorrelated)
detection efficiency for gammas, of the different trigger efficiency 
MiniMax has for
events in which no particles are detected, and possess definite and 
very different values
for pure generic and pure DCC production.\cite{kenk2}
(Note that generalizations involving gamma-gamma
correlations are possible, but depend on a  single detector dependent 
parameter in these cases).

\section{Preliminary Comparison with Monte Carlo Predictions}
Table \ref{tab:data} shows the measured values of lower order ratios 
of the factorial moments.
The results were found to be independent of the run conditions, independent of
the converter thickness or composition, and consistent with a
high $x_F=0.9$ trigger.
The data are consistent with the generic production mechanisms.
Comparison between the different trigger samples, and between the calculated
and Monte Carlo \cite{GEANT} corrected values suggest the errors due to systematic 
effects are in the $5\%$ range.
We believe we are sensitive to admixtures of DCC mechanisms to the $10-20\%$
level, however more work is needed to place a limit on the quantity of
DCC we may observe. These values are dependent on the model used to mix
DCC into the generic fragmentation.

\begin{table}[t]\caption{Values of $r_{ij}$ from the data and Monte 
Carlo\label{tab:data}}
\vspace{0.4cm}
\begin{center}
\begin{tabular}{|l|c|c|c|c|}
\hline 
	& PYTHIA & pure & DCC 	&		\\
$r_{ij}$	& and	 & DCC	& and	&	Data 	\\
	& GEANT	 & 	& GEANT &		\\ \hline \hline
$r_{11}$& $1.01\pm.02$ & $0.500$ & $0.56\pm.01$ & $0.98\pm.01$ \\
$r_{21}$& $1.02\pm.05$ & $0.333$ & $0.40\pm.03$ & $0.99\pm.02$ \\
$r_{31}$& $1.09\pm.14$ & $0.250$ & $0.34\pm.05$ & $1.03\pm.04$ \\ \hline\hline
\end{tabular} \end{center} \end{table}

\section*{Acknowledgments} 
This work has been supported in part by the U.S. Department of Energy, the
National Science Foundation, the Gugenheim Foundation, the Timken
Foundation, and the Case Western Reserve University Provost's Fund.


\section*{References}
\begin{thebibliography}{99} 
\bibitem{raja}C.M.G.Lattes, Y.Fujimoto and S.Hasegawa,  {\em Phys. Rep.}
{\bf 65} 151 (1980).
\bibitem{ansl}For a review, K.Rajagopal, in Quark-Gluon Plasma 2, ed. R.Hwa,
World Scientific, 1995, HUTP-95-A013.
\bibitem{kenk}T.Brooks {\it et al}, The MiniMax Detector.
NIM article in preparation.
\bibitem{kenk2}T.Brooks {\it et al}, Analysis of Charged-Particle/Photon
Correlations in Hadronic Multiparticle Production.
In preparation.
\bibitem{GEANT}GEANT Detector and Simulation Tool, CERN, PM0062 (1993).\\
T. Sjostrand {\em CERN-TH.6488/92 (PYTHIA 5.6 Manual)}
\end{thebibliography}
\end{document}